# Are Digital Humanities really committed to open? An exploratory study on the availability of methodological workflows and open peer review practices

Silvio Peroni

Digital Humanities Advanced Research Centre (/DH.arc)
Department of Classical Philology and Italian Studies, University of Bologna, Bologna, Italy
silvio.peroni@unibo.it - https://orcid.org/0000-0003-0530-4305

**ABSTRACT (ENGLISH)**
Open Science has become a central framework for promoting transparency, accessibility, and inclusiveness in scholarly research. While the Digital Humanities (DH) community has long embraced openness in terms of research outputs, less attention seems to have been paid to the openness of the methodological and evaluative processes underlying knowledge production. This paper presents an exploratory study that investigates the current state of openness in DH research practices, focusing specifically on research data management documentation and peer review processes. In particular, this study addresses two research questions: (1) to what extent DH publications that describe data explicitly reference external documentation detailing data creation and management processes; and (2) how widely open peer review practices are adopted across DH conferences and journals. The results revealed a limited adoption of open methodological practices. Only a small fraction of the analysed articles provided explicit, reusable documentation of data creation workflows, and no references to data management plans or formal research data management documentation were found. An even more critical picture emerges from the analysis of peer review practices: the vast majority of DH venues continue to rely on traditional single- or double-blind review models, with open peer review adopted in only a few isolated cases.

**ABSTRACT (ITALIANO)**
*Le Digital Humanities sono davvero orientate all'open? Uno studio esplorativo sulla disponibilità dei workflow metodologici e sulle pratiche di peer review aperta.* La Scienza Aperta è diventata un quadro di riferimento centrale per promuovere la trasparenza, l'accessibilità e l'inclusività nella ricerca scientifica. Sebbene la comunità delle Digital Humanities (DH) abbia da tempo adottato pratiche di apertura per quanto riguarda i prodotti della ricerca, sembra che sia stata prestata minore attenzione all'apertura dei processi metodologici e valutativi che sottendono la produzione della conoscenza. Questo contributo presenta uno studio esplorativo che indaga lo stato attuale dell'apertura nelle pratiche di ricerca nelle DH, concentrandosi in particolare sulla documentazione relativa alla gestione dei dati della ricerca e sui processi di peer review. In particolare, lo studio affronta due domande di ricerca: (1) in che misura le pubblicazioni DH che descrivono dei dati fanno esplicito riferimento a documentazione esterna che dettagli i processi di creazione e gestione dei dati; e (2) quanto siano diffuse le pratiche di open peer review tra conferenze e riviste DH. I risultati mostrano un'adozione limitata di pratiche metodologiche aperte. Solo una piccola parte degli articoli analizzati fornisce una documentazione esplicita e riutilizzabile dei workflow di creazione dei dati, mentre non è stato riscontrato alcun riferimento a piani di gestione dei dati o a documentazione formale di research data management. Un quadro ancora più critico emerge dall'analisi delle pratiche di peer review: la grande maggioranza delle sedi DH continua ad affidarsi a modelli tradizionali di revisione single-blind o double-blind, mentre le pratiche di open peer review risultano adottate solo in pochi casi isolati.
**Parole chiave:** open science; piano per la gestione dei dati; revisione tra pari aperta; metodologie

## 1. INTRODUCTION

A few years ago, UNESCO (2021) published its guidelines about Open Science, highlighting all the relevant dimensions that interests this "construct" and that involves all scientific disciplines and aspects of scholarly practices in (a) making *scientific knowledge* openly available, accessible and reusable for everyone, and (b) opening the *processes* of scientific knowledge creation, evaluation and communication to societal actors beyond the traditional scientific community. The Digital Humanities (DH) domain has worked for years to open up the scientific knowledge produced by its research, approaching open access as a primary practice across its scholarly venues, from conferences to journals. In addition, more recently, several other initiatives have been launched to promote the publication of non-traditional research objects, such as datasets, software, corpora, and models. For instance, the *Journal of Open Humanities Data* (JOHD, https://openhumanitiesdata.metajnl.com/) has been proposed for publishing peer-reviewed articles describing humanities research objects or techniques with high potential for reuse across all subjects in the Humanities domain. Indeed, the publication of these non-traditional research objects has been demonstrated to have "a positive impact on both the metrics of research papers associated with them and on data reuse" (McGillivray et al., 2022).

From these premises, we can undoubtedly claim that the DH community has done (and continues to do) a lot for supporting the open sharing of scientific knowledge. However, a more blurred picture emerges when we consider the other side of the Open Science endeavour, namely, *how* (rather than *what*) a given research is conducted or evaluated. This aspect relates more to the *methodological process* that governs DH research rather than the outcomes it produces. Indeed, from the one hand, "we need to put more explicit attention on methodologies, and on the need for carefully documenting each step of a research workflow" (Barzaghi et al., 2025), including documenting research data management practices (Tóth-Czifra et al., 2023) via known documentation such as Data Management Plans (DMPs), since data "are of limited use if they are not sufficiently documented, and if they are not accompanied by a clear description of the research methodology" (Barzaghi et al., 2025). On the other hand, we also need to actively work to make research evaluation processes, such as *peer review*, more inclusive and, as a consequence, more transparent (Ross-Hellauer, 2017).

In this paper, we describe an exploratory study to address the aforementioned aspects and begin to understand the current state of DH research by focussing on the openness of its methodological process rather than its research outputs. In particular, we aim at having initial answers to the following research questions (RQ1-2), one related to the *knowledge production* practice and the other to the *knowledge evaluation* procedure:

1. How many publications which describe some data contain an explicit reference to external documentation detailing the process used to create and maintain such data, such as a DMP and/or a detailed workflow depicting the process of data creation?
2. How many DH venues adopt open peer reviewing practices?

To answer these two questions, we have devised two procedural workflows to gather relevant data from two distinct sets of resources. On the one hand, to address RQ1, we have downloaded and analysed, in a semi-automatic fashion, several articles describing data published in JOHD to identify references and mentions to external documentation detailing data management plans and methodological workflows adopted for data creation. On the other hand, to address RQ2, we have analysed several calls for papers and the editorial processes of DH conferences and journals to see how widely open reviewing is adopted within the community.

The rest of the paper is organised as follows. In Section 2, we introduce the materials used and the method developed to gather data to answer the research questions. In Section 3, we present the results obtained by analysing the data. In Section 4, we discuss the results and conclude the paper, sketching out some future work.

## 2. MATERIALS AND METHODS

To address the two RQs, we have developed two distinct workflows, available in (Peroni, 2026a) for reproducibility purposes, which are briefly introduced as follows. In addition, the related data obtained from running the workflow can be found in (Peroni, 2026c), while the related Data Management Plan can be found in (Peroni, 2026b).

**Data Papers from the Journal of Open Humanities Data.** The main goal of this activity is to understand if a specific Digital Humanities venue, i.e. JOHD, includes references or mentions to external material appropriately dedicated to either describing the management of the data a paper describes, such as a Data Management Plan (DMP), or detailing precisely the workflow used to produce these data, e.g. published as a particular workflow document or as a computational notebook. In particular, we focus on specific kinds of articles published in JOHD, i.e. *Data Papers*, which are concise descriptions of a humanities research object with high reuse potential. These articles are particularly suitable for our activities since they describe only data, and thus they should have a higher possibility of having such documents referenced, given that the data are the focus of the article. It is worth clarifying, though, that such Data Papers usually include a methodological section that provides hints about the process of creating the data. However, the information in that section is often insufficient to fully reproduce the workflow steps required to recreate the same set of data.

To have a limited time window for the articles, we consider all the Data Papers included in volume 11 of JOHD, i.e. those published in 2025. We retrieve the basic metadata (DOI and title) for these Data Papers and download the corresponding XML and PDF files. Then, we parse all the XML versions of the articles to find out if their text contains any of the terms usually related with data management and analysis: *data management plan* and its acronym (*dmp*), *research data management* and its acronym (*rdm*), the broad concept of *data management* and, finally, two terms – i.e. *protocol* and *workflow* – that are commonly used to define processes devised for the creation and analysis of data. The rationale for using these terms for filtering relevant articles is that, instead of manually reviewing all downloaded articles, we focus on those that contain words that potentially indicate a mention of external documents defining DMPs, workflows, etc.

Finally, given the list of articles returned in the previous step, we read each article to see if it contains references or mentions to external documents describing the management or the creation of the data it introduces. In particular, we check if either (a) it contains a reference or a mention to such an external document, or (b) such a document is included in the data deposited and introduced in the article.

**Reviewing processes in Digital Humanities venues.** The main goal of this activity is to identify the practice of reviewing processes established in some DH venues, including conferences and journals. In particular, we want to check how many venues adopt open peer review processes instead of the classic ones that prescribe either the sole anonymity of reviewers (*single-blind* from now on) or the anonymity of both authors and reviewers (*double-blind* from now on).

Since the concept of open peer review can be decoupled along different dimensions depending on the possibilities a peer-reviewing process permits, we decided to adopt the taxonomy defined in (Ross-Hellauer, 2017), which comprises the following dimensions:

- open identities: authors and reviewers are aware of each other's identities;
- open reports: review reports are published alongside the relevant article;
- open participation: the wider community is able to contribute to the review process;
- open interaction: direct reciprocal discussion between author(s) and reviewers, and/or between reviewers, is allowed and encouraged;
- open pre-review manuscripts: manuscripts are made immediately available (e.g., via pre-print servers like arXiv) in advance of any formal peer review procedures;
- open final-version commenting: review or commenting on final "version of record" publications;
- open platforms ("decoupled review"): review is facilitated by a different organisational entity than the venue of publication.

We consider the following DH venues, of which websites have been analysed to find information about their call for papers and, in particular, the reviewing process adopted:

- the past five years of the annual conference organised by the *Associazione per l'Informatica Umanistica e la Cultura Digitale* (AIUCD, https://www.aiucd.it/);
- the past five years of the annual conference organised by the *Alliance of Digital Humanities Organizations* (ADHO, https://adho.org/);
- the DH journals in the list created by Spinaci et al. (2022) available at https://dhjournals.github.io/list/, in particular those that are (a) completely devoted to publishing DH articles, (b) still active, (c) have a URL associated, and (d) specify a standard article submission procedure.

All the retrieved information about the peer review process adopted by the various venues is extracted. When tracking open peer review processes, it is possible to consider more than one dimension per venue simultaneously if multiple dimensions are actually considered in the reviewing process. In addition, we also track the venues that do not clarify on their website the review process in use.

### 3. RESULTS

From JOHD, we downloaded 48 Data Papers, all published in 2025. From these, 17 articles matched at least one of the searched words, distributed as shown in the left graph in Figure 1. The terms *data management plan*, *DMP*, *research data management*, and *RDM* were not mentioned in any article. Out of these 17 articles, only 5 articles (around the 10% of the total) provided a full description of the protocol followed for creating the data introduced in the article, either as an external document or included in the data deposited in a relevant repository (e.g. Zenodo) [RQ1]. The 3 articles mentioning *data management* did not include any reference/mention to protocols, DMPs, or other relevant documentation.

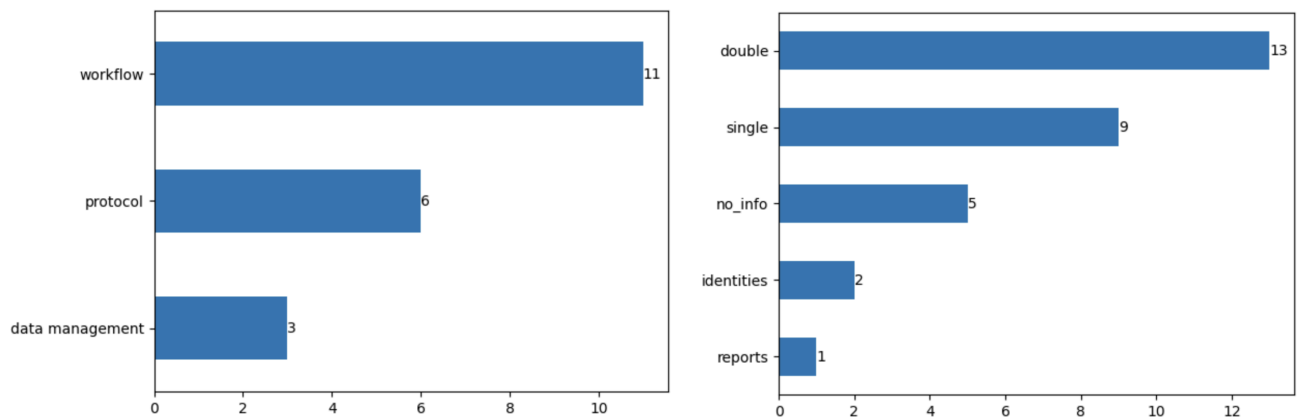

**Figure 1. Two horizontal bar charts depicting some of the results obtained from the analysis. The left graph shows the distribution of the searched terms in the downloaded JOHD Data Papers, where each article may match more than one term. The right graph shows the different peer-reviewing processes adopted by the DH venues analysed, where "no_info" indicates those venues that did not have a clear reviewing process specified. Both graphs report only the dimensions that have been encountered at least once.**

We have considered 29 distinct DH venues: 10 conference editions and 19 journals. The majority of the venues adopted a traditional reviewing process, with a preference for double-blind peer review (13 venues) over the single-blind peer review (9 venues), while 5 venues did not provide sufficient information to understand the peer review procedure used. Overall, only two venues adopted some form of open peer review in the past, i.e. DH 2023 (*open identities*) and DH 2025 (*open identities* and *open reports*) [RQ2]. The 5 AIUCD conference editions analysed, in contrast, had all adopted a double-blind peer review process.

## 4. DISCUSSIONS AND CONCLUSIONS

The results introduced by our exploratory study seem to suggest that there is still work to do by the DH community to properly address the openness of the Open Science area dedicated to the processes that regulate knowledge creation and evaluation. Indeed, looking with more detail at the results obtained by analysing the JOHD articles, it seems peculiar that, in all the Data Papers published in 2025, no mentions of the terms *data management plan* (and *DMP*) and *research data management* (*RDM*) have not been found in any of the 49 articles. While explicitly required by funders, such as the European Commission in all its funding programmes, it seems that the practice of including them as additional documentation to support the data researchers create and/or gather for answering specific questions is not systematically followed, even when looking at a venue, JOHD, explicitly dedicated to introducing data for Humanities research.

While DMPs were completely missing, we observed that a few articles at least provided an appropriate, detailed description of the procedural methodology used to generate the data described in the articles analysed. While not ideal, since it has not been systematically observed across all the articles considered, it seems that the community focuses more on the reproducibility of the approach used to generate the data than on the administrative and managerial aspects of data before and after the research is completed. Another aspect that deserves future analysis in this regard is how detailed data creation workflows are organised to create data that are FAIR-by-design (Wilkinson et al., 2016), a fundamental practice to consider for reliable research data management, as highlighted by UNESCO (2022). In addition, recently several policy documents push for complementing the FAIR Principles with the CARE (*collective benefit*, *authority to control*, *responsibility*, *ethics*) principles (Carroll et al., 2020), which put "people and purpose, rather than data and technology" at the centre of RDM frameworks (Di Donato & Provost, 2025). Also, these aspects should be taken into strong consideration to address transparent and responsible RDM in DH projects.

The analysis of peer review process adoption in the considered venues revealed a worse scenario than the RDM practices mentioned above. With the exception of a few conference events that attempted to address at least the basic requirements of open peer review, the majority of DH venues do not currently support transparent, reliable processes for evaluating the scientific appropriateness of DH research contributions. While there have been ongoing discussions for years on the pros and cons of such a new peer review dynamics (Ross-Hellauer, 2017), all Open Science practices and norms released in the past years strongly support the open peer review process as the primary alternative to enable a broader setting of trust of the research endeavour. The hope is that this topic will spark constructive discussions within the community to understand how it can be implemented as part of modern policies for conducting transparent, reliable research.

**ACKNOWLEDGEMENTS**
This work has been funded by the European Union's Horizon Europe framework program under grant agreement No 101188018 (GRAPHIA).